# Gameplay experience in a gaze interaction game

**Lennart Nacke, Sophie Stellmach, Dennis Sasse, Craig A. Lindley**
Game and Media Arts Laboratory, Department of Interaction and System Design, School of Engineering,
Blekinge Institute of Technology
PO Box 214, SE-374 24 Karlshamn, Sweden
Lennart.Nacke@bth.se, sstellmach@gmail.com, Dennis.Sasse@gmail.com, Craig.Lindley@bth.se

**Keywords**
Game experience, flow, immersion, gaming with gaze, human-computer interaction

## Introduction

Eye tracking has been well researched in the field of human-computer interaction (Duchowski, 2007; Jacob, 1990). In addition, interacting with gaze as an input modality for digital games has recently become a popular area of study (Isokoski et al., 2007; Jönsson, 2005; Kenny et al., 2005; Smith and Graham, 2006; Špakov, 2005).

Jönsson (2005) compared eye and mouse control as input for two three-dimensional (*3D*) computer games and found that gaze control was more accurate, game experience was perceived as subjectively more enjoyable and committing. Smith and Graham (2006) studied eye-based input for several game types, with principally *3D* navigation. Their results show that participants felt more immersed when using the eye tracker as a gaming input device. Kenny et al. (2005) developed a first-person shooter (*FPS*) game that logs eye tracking data, video data and game internal data, which were correlated with each other. They found that players fixate the center of the screen for a majority of the time. Isokoski et al. (2007; 2006) describe the advantage of gaze pointing in *FPS* games as aligning the camera (view frustum of the player) to the target becomes obsolete, when aiming is decoupled from view. Their results indicate that gaze input for *FPS* games can compete with "killing efficiency" of gamepad input, but leads to more ammunition used due to problems of targeting accuracy. Thus, the game developed for this study was focusing solely on navigational challenges using gaze as input to a *3D FPS*, allowing us to assess *gameplay* experience with subjective questionnaires for this novel input modality.

In contrast to previous studies, which mainly compared mouse and gaze interaction in terms of efficiency and accuracy (Agustin et al., 2007; Dorr et al., 2007; Isokoski et al., 2007), the purpose of the study reported here is to broadly investigate gameplay experience in a *3D* gaze interaction game by testing the reliability of a range of subjective experiential questionnaires. As there are currently no established measures for assessing notions of, for example, *flow* (Csikszentmihalyi, 1990) in games, this investigation may be the foundation for a more thorough comparison of input modalities using subjective experience assessment questionnaires. While gameplay experience itself is currently not well understood, this is especially true for gaze interaction games. Thus, Gowases et al. (2008) claimed that it is very important to evaluate immersion and user experience in eye tracking games. Their study shows that players felt more immersed when using a gaze-based input method in comparison to using a mouse. We conducted a field experiment to investigate gaze steering in a *3D* game[1]. We assess gameplay experience in detail for this gaze interaction game using self-report game experience (IJsselsteijn et al., n.d.), flow (Csikszentmihalyi, 1990) and presence (Vorderer et al., 2004) questionnaires. The novelty of this paper stems from it using statistical correlations of the previously mentioned experience questionnaires to explore key points of game experience in a gaze interaction game.

---

[1] A video of the field experiment game stimulus and setup can be watched online at http://www.youtube.com/watch?v=6PZpsWzjnvE

## Method

**Stimulus.** The stimulus for this experiment was a game-mod level developed using the *Half-Life 2* Source SDK platform (Valve Corporation, 2004). The game level was designed around the concept of navigation in a *3D* virtual environment, thus, the goal was to navigate successfully (i.e. without falling off) on a catwalk to a door indicating the end of the level (see Figure 1). The keyboard keys (*W-A-S-D*) were used to control locomotion of the player, while gaze input was used to control the first-person camera view[2] (usually controlled with the mouse in *FPS*). Navigational challenges were the labyrinthine structure of the catwalk and several oil drums placed in the way of the player. Players highscore was evaluated according to playing time and times they fell off the catwalk. We chose to use this game, since we wanted to investigate the gameplay experience related to navigational challenge of using gaze as a steering method in an *FPS* game.

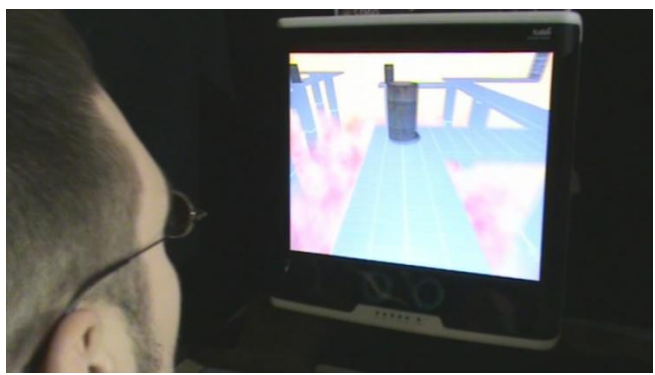

**Figure 1.** The 3D gaze interaction stimulus, a *Half-Life 2* level.

**Participants.** Data were recorded from *30* random people, *2* female and *28* male, at *Dreamhack Winter 2007*[3], aged between *14* and *32* (*Mean (M)* = 18.67, *Standard Deviation (SD)* = 4.26). Participants were gamers that were interviewed at a booth at *Dreamhack* and invited to participate in the experiment. *76.7%* started playing games when they were younger than *12* years old, *23.3%* did this when they were between *12* and *20* years old. *53.3%* considered themselves to be skilled players. *86.7%* preferred keyboard and mouse as their gaming input device, while the rest opted for joypad and joystick. Four people wore glasses and four people wore contact lenses (all of them near-sighted), one person was a bit cross-eyed with the left eye pointing a bit more to the left. Finally, *56.7%* indicated that they play computer games every day.

**Experimental design and apparatus.** The experiment was designed to assess gaze steering experience in a game level. Our questionnaires only focused on evaluating gaze-input game experience. However, we did allow participants to additionally play the game levels with a mouse[4]. Each participant was allowed to get used to playing with gaze in another test level first and then played the stimulus level until they had reached the door at the end of the level. Components of game experience were measured for the gaze condition using a game experience questionnaire (*GEQ*) (IJsselsteijn et al., n.d.), which measures the experiential dimensions of *immersion*, *tension*, *competence*, *flow*, *negative affect*, *positive affect*, and *challenge*. Each component consists of *6* items – each indicating a feeling statement – to which agreement is measured on a five-point scale ranging from *0* (not agreeing with the statement) to *4* (completely agreeing), thus resulting in *42* statements. Component scores are computed as the average value of its items. The questionnaire is based on focus group research (Poels et al., 2007) and subsequent survey investigations among frequent players as part of the *EU*-funded *FUGA* project (Contract: *FP6-NEST-28765*). In addition to that the Flow State Scale (*FSS*) from Jackson and Marsh (1996) and items from the *MEC* Spatial Presence Questionnaire (Vorderer et al., 2004) were used to check for possible correlations with the *GEQ*. The gaze game was played using a *Tobii T120* eye tracker.

---

[2] The mapping of gaze input can be described like this: If a player looks to the left side of the screen, the camera (or view) will also turn to the left, if he looks to the right side, the view will turn right, etc. - player controls her view in the virtual world by looking at different areas of the monitor.

[3] Dreamhack is the world's largest computer game festival, for more information see http://www.dreamhack.se

[4] Again, the primary focus of this study was not a comparison between gaze and mouse input, but an assessment of game experience for gaze interaction. The playing times with the mouse were much shorter (~1 minute), thus not resulting in measurable experience. We cross-checked the difference between gaze and mouse input with a few questions: *83.4%* indicated that they had to concentrate more for navigating correctly with gaze interaction. *63.3%* preferred gaze input to mouse input as interaction method. Only, *43.3%* felt limited by the gaze interaction in their navigation through the level.

**Procedure.** The experiment was conducted as part of a booth from Blekinge Institute of Technology that was set up at *Dreamhack Winter* in Jönköping, Sweden from November 29 until December 2, 2007. Participants were selected from visitors of the event and invited to participate in the study to enter a lottery for winning a radio-controlled mini-helicopter at the end of the event. After playing, participants were asked to fill out the questionnaires, then thanked for their participation and escorted off the booth.

## Results

The results of the game experience questionnaire showed a very positive game experience. Most notably positive affect ($M = 3.31$, $SD = 0.45$), immersion ($M = 3.11$, $SD = 0.57$) and flow ($M = 2.84$, $SD = 0.74$) scored high for the gaze input game. Challenge ($M = 2.63$, $SD = 0.52$) and competence ($M = 2.37$, $SD = 0.80$) scored high compared to prior studies (Grimshaw et al., 2008; Nacke and Lindley, 2008). As expected, negative affect ($M = 0.37$, $SD = 0.41$) and tension ($M = 1.01$, $SD = 0.52$) dimensions scored low, indicating a very pleasant game experience. Figure 2 shows average gameplay experience scores of the *GEQ* for the gaze interaction game.

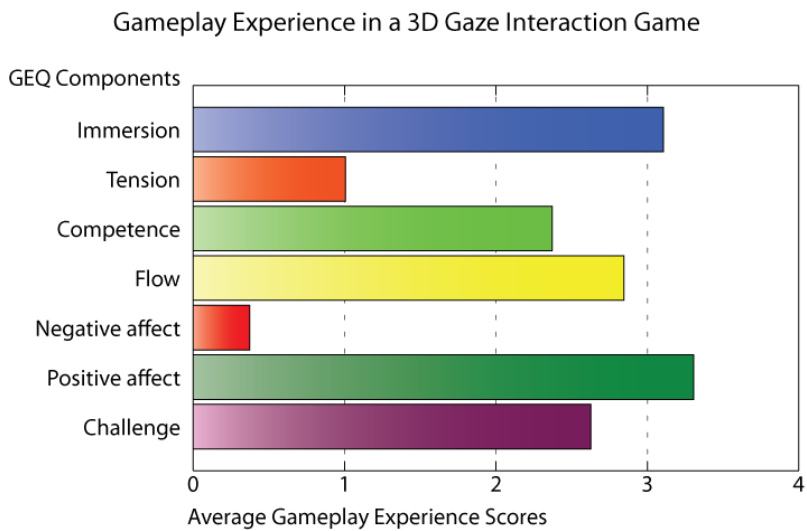

**Figure 2.** Average gameplay experience component scores recorded from playing the gaze interaction *Half-Life 2* mod.

The *FSS* (Jackson and Marsh, 1996) rates *36* items in a *1* (strongly disagree) to *5* (strongly agree) response format to assess *flow* experience (Csikszentmihalyi, 1990). It consists of several subscales (see Figure 3). The average flow value – calculated as a mean of all items – was high ($M = 3.59$, $SD = 0.55$) in comparison to values reported by Jackson and Marsh (1996). The highest flow component was *autotelic experience* ($M = 4.43$, $SD = 0.77$). Figure 3 shows the average scores of all the subscales in comparison.

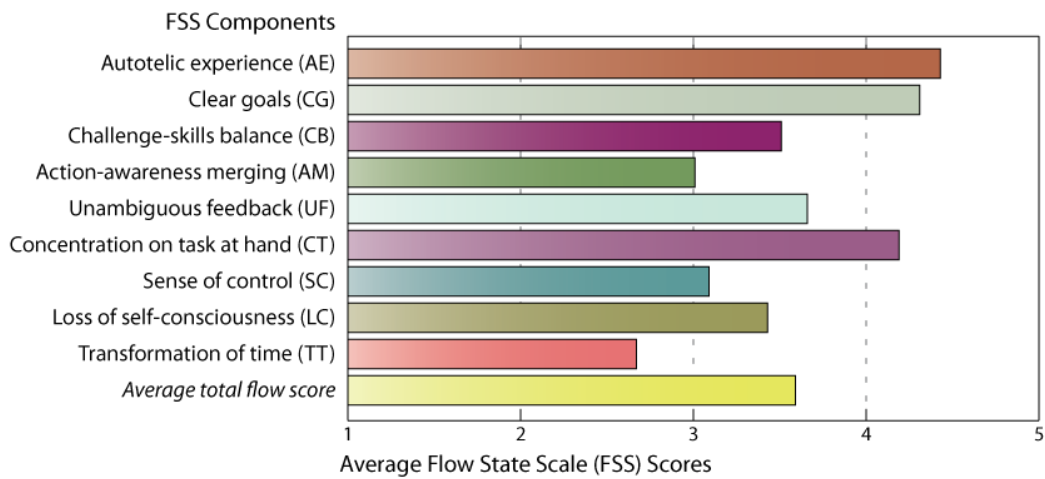

**Figure 3.** Average *FSS* scores recorded from playing the gaze interaction *Half-Life 2* mod.

Finally, we assessed mean scores for the *MEC* Spatial Presence Questionnaire (Vorderer et al., 2004). It can be noted that spatial presence possible actions ($M = 3.68$, $SD = 0.81$) ratings were significantly higher than spatial presence self-location ($M = 3.21$, $SD = 0.02$) ratings, $t(29) = -3.22$, $p < .01$. This could be attributed to the gaze interaction experience being more immersive than the content in the game.

When correlating the questionnaire items between *GEQ* and *FSS*, we found highly significant positive correlations (using Pearson's $r$[5]) between *challenge-skills balance* and *competence* ($r = .57^{*}$[6]), *clear goals* and *competence* ($r = .49^{*}$), *sense of control* and *competence* ($r = .61^{*}$), *autotelic experience* and *competence* ($r = .55^{*}$) and the overall *FSS flow* value and *GEQ competence* ($r = .58^{*}$). The *FSS flow* value and the *GEQ flow* value did not significantly correlate, in fact they were significantly different, $t(29) = -4.41$, $p < .001$. Interestingly, we also found very significant positive correlations between the items *unambiguous feedback* and *positive affect* ($r = .48^{*}$), *transformation of time* and *spatial presence self location* ($r = .50^{*}$), and between *autotelic experience* and *immersion* ($r = .51^{*}$). Also of interest was a significant *negative* relationship between *sense of control* and *challenge* ($r = -.37$, $p < .05$), indicating that less interaction control results in higher challenges. These relationships indicate the complexity of game experience for *3D* gaze interaction games.

# Conclusion

Due to the experimental setting, the novelty of navigating in a game at a computer festival could be high. However, our findings indicate that gaze interaction games[7] provide a positive game experience, where the challenge of controlling the game by gaze (and keyboard) results in positive affection and feelings of flow and immersion, which can be reliably assessed using the questionnaires presented in this study. In line with the results of Gowases et al. (2008), *immersion* and *spatial presence possible actions* ratings for the gaze interaction game focusing on navigational challenge, were high and a relationship between *autotelic experience* and *immersion* was found, which demands further investigation. Having assessed questionnaires as usable tools to investigate game experience with a gaze navigation game, we plan to conduct more studies in a controlled laboratory setting that will allow us to study and log (see Nacke et al., 2008) different game and input types more thoroughly.

---

[5] Pearson's correlation coefficient ($r$) is a standard measure for testing the relationship strength between two variables.

[6] * denotes $p < .01$ (which stands for highly significant correlations in this case)

[7] In this context, gaze interaction games refers to *3D* gaze interaction *FPS* similar to the mod that we have created, where the player exerts effort to control her view in the virtual world by looking at different areas of the monitor in the real world. This way of controlling camera view in an *FPS* game seems to have exciting potential in terms of gameplay experience and will serve us as an entry point to future experiments.

## Acknowledgements

This research was supported by "*FUGA - The Fun of Gaming: Measuring the Human Experience of Media Enjoyment*", funded by the European Commission (Contract: *FP6-NEST-28765*). We thank all participants involved in this study. Also, we would like to thank Charlotte Sennersten for her help in organizing the Dreamhack experiments as well as Jessica Lundqvist, Anders Vennström and their colleagues from *Tobii Technology AB* for supplying and supporting us with the eye trackers.